\documentclass[twocolumn,superscriptaddress,floatfix,prb,aps,10pt]{revtex4-1}
\usepackage[utf8]{inputenc}
\usepackage{graphicx}
\usepackage{textcomp}
\usepackage{units}
\usepackage{siunitx}
\usepackage{amsmath}
\usepackage{amsfonts}
\usepackage{amssymb}
\usepackage{multirow}
\usepackage{bigstrut}
\usepackage{tabularx}
\usepackage{cancel}
\usepackage{xint}
\usepackage[usenames,dvipsnames,svgnames,table]{xcolor}

\renewcommand{\vec}[1]{\boldsymbol{#1}}
\newcommand{\bra}[1]{\bigl\langle#1\bigr|}
\newcommand{\ket}[1]{\bigl|#1\bigr\rangle}
\newcommand{\bracket}[2]{\bigl\langle#1\big|#2\bigr\rangle}
\newcommand{\matx}[3]{\bigl\langle#1\big|#2\big|#3\bigr\rangle}
\newcommand{\expect}[1]{\bigl\langle #1 \bigr\rangle}

\newcommand{\Ham}{\ensuremath{\mathcal{H}}}
\newcommand{\Hs}[1]{\ensuremath{\mathcal{H}_\mathrm{#1}}}
\newcommand{\HCF}{\Hs{CF}}
\newcommand{\HZ}{\Hs{Z}}
\newcommand{\J}[1]{\ensuremath{\hat{J}_{#1}}}
\newcommand{\Jz}{\J{z}}
\newcommand{\Jp}{\J{+}}
\newcommand{\Jm}{\J{-}}
\newcommand{\Ji}{\J{i}}
\newcommand{\T}{\ensuremath{\mathcal{T}}}  
\newcommand{\R}[1][1]{\ensuremath{\mathcal{R}^{{\xintiiMul{2}{#1}}\pi/q}}}  
\newcommand{\Rx}{\ensuremath{\mathcal{R}^\pi_x}} 
\newcommand{\Sq}{\ensuremath{\mathcal{S}^{\pi/q}}}  
\newcommand{\Rhalf}{\ensuremath{\mathcal{R}^{\pi/q}}}  

\newcommand{\mathcomma}{\,,}
\newcommand{\mathperiod}{\,.}

\makeatletter
\def\CT@@do@color{%
    \global\let\CT@do@color\relax
        \@tempdima\wd\z@
        \advance\@tempdima\@tempdimb
        \advance\@tempdima\@tempdimc
\advance\@tempdimb\tabcolsep
\advance\@tempdimc\tabcolsep
\advance\@tempdima2\tabcolsep
        \kern-\@tempdimb
        \leaders\vrule
                \hskip\@tempdima\@plus  1fill
        \kern-\@tempdimc
        \hskip-\wd\z@ \@plus -1fill }
\makeatother

\begin{document}

\title{Electron-assisted magnetization tunneling in single spin systems}

\author{Timofey Balashov}
\affiliation{Physikalisches Institut, Karlsruhe Institute of Technology, Wolfgang-Gaede-Strasse 1, 76131 Karlsruhe, Germany}
\author{Christian Karlewski}
\affiliation{Institut f\"ur Theoretische Festk\"orperphysik, Karlsruhe Institute of Technology, Wolfgang-Gaede-Strasse 1, 76131 Karlsruhe, Germany}
\affiliation{Institute of Nanotechnology, Karlsruhe Institute of Technology,  76344 Eggenstein-Leopoldshafen, Germany}
\author{Tobias M\"arkl}
\affiliation{Physikalisches Institut, Karlsruhe Institute of Technology, Wolfgang-Gaede-Strasse 1, 76131 Karlsruhe, Germany}
\affiliation{University of Canterbury, Dept. of Physics and Astronomy, Christchurch 8140, New Zealand}
\author{Gerd Sch\"on}
\affiliation{Institut f\"ur Theoretische Festk\"orperphysik, Karlsruhe Institute of Technology, Wolfgang-Gaede-Strasse 1, 76131 Karlsruhe, Germany}
\affiliation{Institute of Nanotechnology, Karlsruhe Institute of Technology,  76344 Eggenstein-Leopoldshafen, Germany}
\author{Wulf Wulfhekel}
\affiliation{Physikalisches Institut, Karlsruhe Institute of Technology, Wolfgang-Gaede-Strasse 1, 76131 Karlsruhe, Germany}
\affiliation{Institute of Nanotechnology, Karlsruhe Institute of Technology,  76344 Eggenstein-Leopoldshafen, Germany}

\begin{abstract}
Magnetic excitations of single atoms on surfaces have been widely studied experimentally in the past decade. Lately, systems with unprecedented magnetic stability started to emerge. Here, we present a general theoretical investigation of the stability of rare-earth magnetic atoms exposed to crystal or ligand fields of various symmetry and to exchange scattering with an electron bath. By analyzing the properties of the atomic wavefunction, we show that certain combinations of symmetry and total angular momentum are inherently stable against first or even higher order interactions with electrons. Further, we investigate the effect of an external magnetic field on the magnetic stability.
\end{abstract}

\date{\today}

\maketitle
\setlength{\parskip}{0cm}

\section{Introduction}
The experimental discovery of large magnetic anisotropy in single Co atoms on a Pt(111) surface~\cite{Gambardella2003} has given hope that systems as small as one atom can be used for information storage in real-life applications. The magnetic properties of single adsorbed atoms have since been investigated for a large range of atom-substrate combinations~\cite{Heinrich2004,Hirjibehedin2007,Balashov2009,Loth_Nature_Physics2010,Loth_New_Journal2010,Loth_Science2010,Lorente2010,Schuh2012,Miyamachi2013,Donati2014,Natterer2017,Baltic2016} using low-temperature scanning tunneling microscopy (STM) and X-ray magnetic circular dichroism (XMCD). Both techniques allow to obtain information on the magnetic anisotropy, i.e.\ the zero field splitting (ZFS) of the magnetic states, caused by the crystal field, and the magnetization decay, i.e.\ the lifetime $T_1$, caused by magnetization tunneling~\cite{Abragam1970}.

In most of the cases reported, the lifetime of the magnetic state was found to be short, with the atom switching between two degenerate antiparallel configurations on a timescale of pico- to microseconds. The interaction of the atom with conduction electrons of the substrate is an important effect limiting the stability of these states~\cite{Balashov2009,Loth_Nature_Physics2010,Loth_New_Journal2010,Loth_Science2010,Lorente2010,Schuh2012}. Two ways of reducing the interaction were proposed: one by physically separating the atom and the conducting substrate by a thin insulator~\cite{Heinrich2004,Hirjibehedin2007}, the other by using rare-earth atoms, where localized 4$f$ electrons, responsible for the magnetic moment, are shielded from the substrate electrons~\cite{Schuh2012}. Both approaches extended the magnetic lifetimes. Lifetimes to the order of minutes at zero magnetic field were found for Holmium atoms on Pt(111)~\cite{Miyamachi2013} followed by an even higher stability in Ho atoms on MgO~\cite{Donati2016,Natterer2017}, persisting in the presence of magnetic field. Dy atoms on graphene on Ir were also found to be stable up to \SI{1000}{\second}~\cite{Baltic2016}. In the latter three reports, the role of the symmetry of the crystal field for stability was highlighted.

In this work we focus on the influence of symmetries on magnetic stability and provide a theoretical framework of stability criteria for rare-earth systems that follow logically from the combination of symmetries of the magnetic atom and its environment. We start with the analysis of $C_q$ point group ($q$-fold rotational symmetry) as the one relevant for atoms adsorbed on a crystal surface, and later extend the results to $C_{qv}, C_{qh}, D_q, D_{qh}, D_{qd}$ and $S_{2q}$ point groups. The results are thus applicable not only to single magnetic atoms on surfaces but also to larger complexes (e.g.\ magnetic molecules) containing a single spin. One can also imagine removing the limitation of rare-earth atoms by considering an effective spin. Further, we do not consider phonon-mediated processes due to their ill-defined angular momentum and for the moment neglect the role of the nuclear spin.

During the preparation of this manuscript several other theoretical works analysing magnetic stability have been published~\cite{Huebner2014, Marciani2017, Prada2017}. This work, however, allows stability analysis without restricting the ground state spin or the magnetic field direction. Further, we extend the stability analysis to higher-order processes and higher symmetries. Interestingly, unconstrained analysis gives different predictions about stability, although identical Hamiltonians are studied. We discuss the differences in the results at the end of the paper.

\section{Symmetries of the Hamiltonian}
\label{symmetries}
A magnetic atom in an anisotropic environment can be generally described by the following Hamiltonian:
\begin{equation}
\mathcal{H} = \mathcal{H}_\mathrm{el} + \mathcal{H}_\mathrm{SO} + \HCF,
\end{equation}
where $\mathcal{H}_\mathrm{el}$ describes the Coulomb interactions within the atom, $\mathcal{H}_\mathrm{SO}$ is the spin-orbit interaction and {\HCF} is the crystal field, i.e.\ the electrostatic interaction of the atom with the environment. For rare earth atoms, the spin-orbit coupling is generally stronger than the effect of the crystal field. Therefore, the last term can be treated as a perturbation on the free atom's lowest $J$ multiplet. We can denote the components of the multiplet as $\ket{L,S,J,m}\equiv\ket{m}$ with fixed values for $L,S$ and $J$ and $m = -J, -J+1, \hdots, J$. We choose the quantization axis $z$ along the highest order rotation axis, i.e.\ perpendicular to the surface when atom on a surface is considered. Note that for transition metals, the crystal field usually exceeds the spin-orbit interaction leading to the partial or full quenching of the orbital momentum. Further, the atomic states can be a complex mixture of several electronic configurations. Thus, often an effective spin is used to describe the low energy part of the atomic states \cite{Heinrich2004,Hirjibehedin2007}. In this case, care has to be taken when analyzing magnetic stability.

As has already been reported~\cite{Miyamachi2013, Marciani2017, Prada2017}, two inherent symmetries of {\HCF} play an important role for the properties of the atom wavefunction: the rotational symmetry and the time-inversion symmetry.

\textbf{Time-inversion symmetry:} The Hamiltonian {\HCF} is due to the electrostatic interactions and thus {\HCF} has to commute with the time reversal operator \T (see~\cite{bookGT, Marciani2017, Prada2017}). {\T} is antiunitary, i.e.\ for any states $\ket{a}$ and $\ket{b}$, $\bracket{a}{b}=\bracket{\T b}{\T a}$. {\T} further anticommutes with all components of the total angular momentum $\vec{J}$, such that $\T\vec{J} = -\vec{J}\T$ and $\T\hat{J}_\pm = -\hat{J}_\mp \T$. The angular momentum eigenstates transform according to $\T\ket{m} = (-1)^{m}\ket{-m}$ which implies $\T^2 = 1$ for integer $J$ and $\T^2 = -1$ for half-integer $J$ (see e.g.\ Refs.~\onlinecite{bookGT, bookQM}). Since {\HCF} commutes with \T, for any eigenstate $\ket{\Psi}$ also the time-reversed state $\T\ket{\Psi}$ is an eigenstate of {\HCF} with the same energy. As time reversal also changes the sign of the expectation value of {\Jz} of such a state, we conclude that the states with non-zero $\expect{\Jz}$ form degenerate pairs~\footnote{As $\ket{\Psi}$ and $\T\ket{\Psi}$ are in this case both eigenstates with the same energy and different $\expect{\Jz}$, we conclude that two orthogonal eigenvectors can be constructed from them.}. If singlet states exist, their $\expect{\Jz}$ is zero, i.e.\ they are effectively non-magnetic and do not change under time reversal (modulo a phase). Such states are only possible in systems with integer $J$, as for half-integer $J$ the spectrum consists exclusively of Kramers' doublets.
Since we are interested in magnetization tunneling, we only consider doublet states in the following. Higher order degeneracies are only possible with fine-tuning the parameters of {\HCF}, which we explicitly exclude~\footnote{The irreducible representations of $C_n, D_n, \text{ and } S_{2n}$ point groups have maximal order 2, so the highest possible degeneracy order is a doublet, barring accidental degeneracy.}. We note that singlet states also usally come in pairs, as symmetric and antisymmetric combinations of two magnetic states with opposite $\expect{\Jz}$, and additional energy terms might reverse singlet formation. We will consider this case later in the section on the influence of magnetic field.

\textbf{Rotational symmetry:} In $C_q$ symmetry, the environment of the magnetic atom is invariant under a rotation by $2\pi/q$, and so {\HCF} also commutes with the rotation operator \R (see~\cite{bookGT, Marciani2017, Prada2017}). Therefore, we can choose the eigenstates of the system to be eigenstates of both {\HCF} and \R. Such eigenstates are superpositions of $\ket{m + nq}$ kets, as all such kets acquire the same phase $\phi = 2\pi m/q$ under rotation.

In the following we use the example of $C_5$ symmetry (Fig.~\ref{fig_c5}), although it is not found on crystal surfaces. Five-fold rotational symmetry does, however, occur in magnetic molecules, and has recently attracted interest regarding magnetic stability~\cite{King2016, Goodwin2017}. Most importanly, it is the lowest symmetry that demonstrates all the relevant effects. A more practically-minded reader can use Fig.~\ref{fig_c_all} as a reference, that also contains examples of all effects, spread over more practically relevant systems.
In the aforementioned figures the $q$ rotation classes are visualized as points on a circle in the complex plane (Fig.~\ref{fig_c5}a \& b). The points represent the phases $\phi$ gained by the application of the rotation operator. A given eigenstate belongs to one and only one class.

Note that for states $\ket{a}$ and $\ket{b}$ belonging to different classes, $\matx{a}{\Jz}{b}$ is zero. This follows from the fact that {\R} and {\Jz} commute. The same relation can be achieved for two degenerate states from the same class by choosing the corresponding eigenvector basis. This choice removes the ambiguity in the definition of eigenstates and maximizes $|\expect{\Jz}|$ for each state~\footnote{We consider the $2\times2$ matrix representation of {\Jz} in the doublet space. Since $\Jz$ is a Hermitian operator, the off-diagonal matrix elements $o$ and $o^*$ are complex conjugate to each other and the diagonal matrix elements have the same magnitude $d$ but different signs due to time inversion symmetry. The determinant of this matrix is thus $-d^2-|o|^2$ and its value cannot depend on the choice of basis. Thus, for $o=0$, $d$ will reach its maximum.}, mimicking the interaction between the single spin and the electron bath~\cite{Delgado2015}.

Such a choice of eigenstates has one more important consequence. Since an eigenstate $\ket{\Psi}$ belongs to a particular class and consists of a well defined set of $\ket{m}$ basis states, its time reversed counterpart $\T\ket{\Psi}$ consists of $\ket{-m}$ basis states and thus belongs to a class with a reversed sign of $\phi$ (which might be the same class). On the class diagram (Fig.~\ref{fig_c5}a), these two classes are symmetric about the $x$ axis. If the two classes are different, $\ket{\Psi}$ and $\T\ket{\Psi}$ are automatically orthogonal, and thus $\T\ket{\Psi}$ must be the other state of the doublet up to a phase factor. We can write for two states $\ket{\Psi^+}$ and $\ket{\Psi^-}$ belonging to the same doublet $\T \ket{\Psi^+} = e^{i\chi}\ket{\Psi^-}$. The phase $\chi$ may be different in every doublet.

\section{Magnetization tunneling}

In an isolated system, any eigenstate is indefinitely stable. For atoms adsorbed on the surface, however, the magnetic atom becomes an open quantum system via the interaction with the substrate.
The magnetization can become unstable if we consider the interaction between the local 4$f$ moment and the bath electrons. The scattering of these electrons on the atom within the $J$ multiplet can be described by the operator $\mathcal{V}$~\cite{Hirjibehedin2007,Fransson2009,Loth_Nature_Physics2010,Loth_New_Journal2010,Delgado2010}.
\begin{equation}
\mathcal{V} = \vec{J} \cdot \vec{\sigma} = \Jz\sigma_z + \frac{1}{2} \left( \Jp\sigma_- + \Jm\sigma_+ \right)\mathcomma
\end{equation}
\noindent
acting in the product space of the atom and the scattering electron ($\vec{\sigma}$ is the spin of the latter). We note that the standard exchange coupling would include the magnetic atom's spin instead of the total angular momentum, but the Wigner-Eckart theorem ensures that $\vec{S}$ can be written as $\vec{J}$ up to a proportionality factor.

This interaction couples different eigenstates of the atom. A system prepared in one of the two ground states can undergo a complete magnetization reversal, i.e.\ electron-assisted magnetization tunneling to the other ground state. We can characterize the stability of the system by the value of the matrix element $V_n = \bra{\Psi_g^-}\mathcal{V}^n\ket{\Psi_g^+}$, where $\Psi_g^\pm$ are the two ground states of the system (remember that we only consider cases where the ground states belong to a doublet). Namely, if $V_1$ is nonzero, a single electron scattering event may reverse the magnetization of the atom. The energy cost of this process is zero, so that the switching will take place at arbitrary low temperatures and would eventually lead to Kondo screening~\cite{Lorente2009}. If $V_1$ is zero, one can consider higher order matrix elements, corresponding to interactions with several electrons.

Since $\mathcal{V}$ operates in a product space of the atom and the interacting electrons, $V_n$ are operators in the electron spin space. The equality $V_n = 0$ is to be understood as every component of it in the $J$ space being equal to zero. Thus, we will consider matrix elements of the form $\matx{\Psi_g^-}{\J{i_n}\dots\J{i_1}}{\Psi_g^+}$, with $i_k =x,y,z$ or, equivalently $i_k = +, -, z$. If all such matrix elements are zero, $V_n$ is a zero operator. Note that such matrix elements describe coherent multi-electron interactions, where the intermediate states are virtual states. This is analogous to co-tunneling~\cite{Averin1990}. Thus, this analysis describes zero-energy transitions of the magnetic state of the atom. At finite temperature we additionally expect incoherent multi-electron interactions, where intermediate states are thermally populated, i.e.\ there is energy transfer between the atom and the bath. This is in analogy to sequential tunneling. In that case the relevant matrix elements are of the form
$\matx{\Psi_g^-}{\J{i_n}}{\Psi_{i_{n-1}}}\dots\matx{\Psi_{i_2}}{\J{i_2}}{\Psi_{i_1}}\matx{\Psi_{i_1}}{\J{i_1}}{\Psi_g^+}$
and a full solution requires solving the complete master equation \cite{Huebner2014,Karlewski2015}.

\begin{figure}
\includegraphics{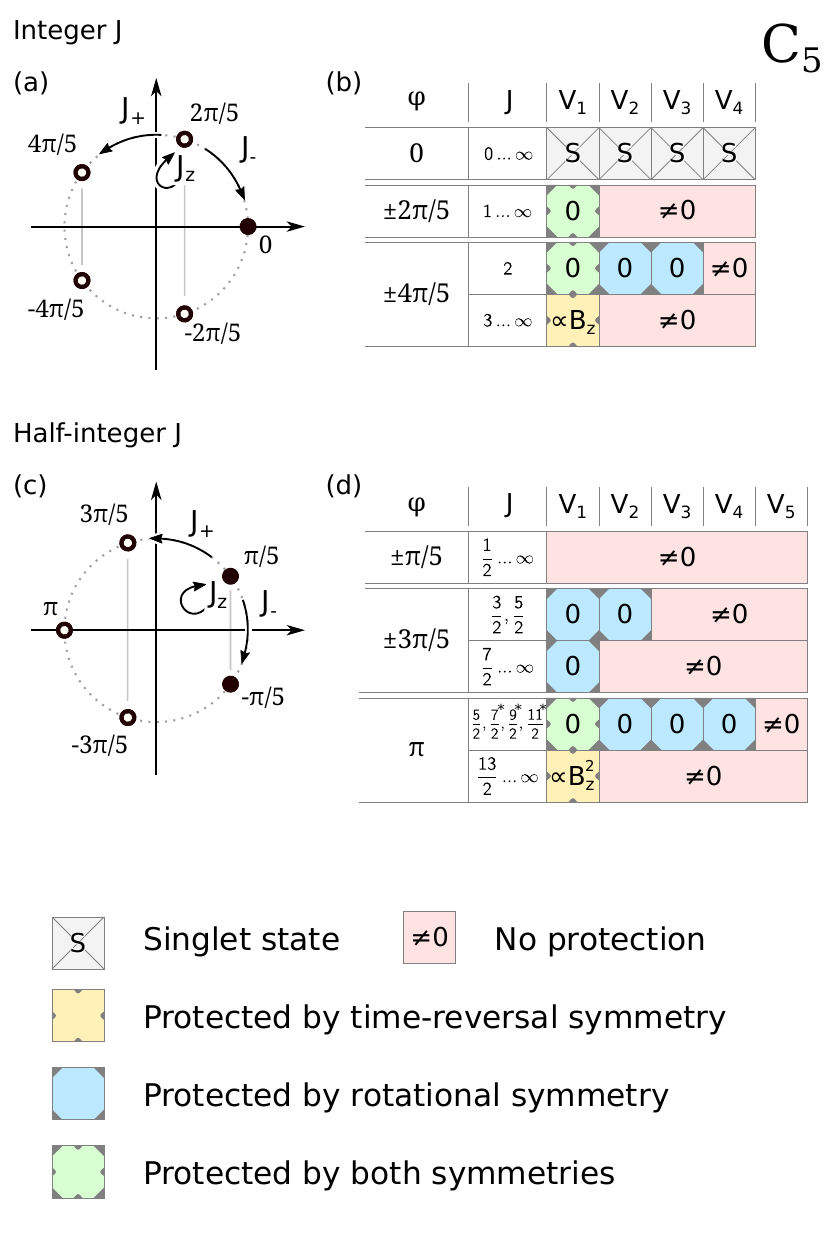}
\caption{Classes and transitions in 5-fold rotation-symmetric environments. (a, c) Rotational classes for integer and half-integer $J$, respectively. Each circle indicates a set of $\ket{m}$ acquiring the same phase under rotation by $2\pi/5$. Empty circles show classes protected against first-order transitions. (b, d) Transition matrix elements between states belonging to particular doublets. Asterisks at certain $J$ values indicate that higher order protection is valid only for coherent electron scattering.}
\label{fig_c5}
\end{figure}

As will be shown, the symmetries of the system lead to $V_1$ being zero for particular classes, thus protecting the ground state doublet from magnetization tunneling.

\textbf{Protection by rotational symmetry:} Let us consider a ground state $\ket{\Psi_g^+}$ from class $2\pi g/q$ and the state $\mathcal{V}\ket{\Psi_g^+}$. The three components of $\mathcal{V}$, namely \Jz, {\Jp} and {\Jm} affect the class of the state differently (see arrows in Figure \ref{fig_c5}a and \ref{fig_c5}b). $\mathcal{V}\ket{\Psi_g^+}$ is a linear combination of eigenstates with classes $2\pi g/q$ (from $\Jz\ket{\Psi_g^+}$), $2\pi (g+1)/q$ (from $\Jp\ket{\Psi_g^+}$) and $2\pi (g-1)/q$ (from $\Jm\ket{\Psi_g^+}$). The other ground state $\ket{\Psi_g^-}$ is from class $-2\pi g/q$. It is evident that $V_1 = \matx{\Psi_g^-}{\mathcal{V}}{\Psi_g^+}$ will be exactly zero unless $-g$ is equal to either $g \text{ or } g\pm1$. This means that the matrix element will vanish when the start and end classes are not neighbors on the circle in the complex plane, or equivalently a single scattering event with a conduction electron can only transfer $\pm1\hbar$. 
For example consider a doublet $\pm3\pi/5$ from Fig.~\ref{fig_c5}c. $\mathcal{V}\ket{\Psi^{\frac35\pi}}$ will only have components from $\pi/5, 3\pi/5 \text{ and } \pi$, and so $\matx{\Psi^{-\frac35\pi}}{\mathcal{V}}{\Psi^{\frac35\pi}} = 0$, as indicated in the table in Fig.~\ref{fig_c5}d by a zero on a blue background. This is the protection mechanism reported in XMCD studies for Ho atoms on MgO and for Dy atoms on graphene~\cite{Donati2014,Baltic2016}.

For higher-order electron scattering processes, that can transfer more than one $\hbar$, the number of classes in $\mathcal{V}^n\ket{\Psi_g^+}$ increases and at a certain order, depending on the separation between the doublet classes, $V_n$ ceases to be zero and zero-energy transitions involving $n$ electrons become allowed. A prominent example of this is the $\pm4\pi/5$ doublet in Fig.~\ref{fig_c5}a for $J=2$. In general, for $2J+1\leq q$ there are at least as many classes as basis states and the states become eigenstates of \Jz. Thus for $J=2$ there is no direct path between $4\pi/5$ and $-4\pi/5$ (i.e.\ $\Jp\ket{2} = 0$). The shortest path in this case contains four transitions between classes, and so $V_4$ is the first matrix element that is non-zero.

A similar situation is realized in the $\pi$ doublet in Fig.~\ref{fig_c5}b for $J=5/2$, as again the two ground states are just $\ket{5/2}$ and $\ket{-5/2}$, and at least five transitions are needed to switch between the two states. Note that the two states in this doublet remain eigenstates of {\Jz} also for higher values of $J$, up to $\frac{11}2$. This means that every intermediate state $\J{i_1}\dots \J{i_n}\ket{\Psi_g^+}$ is also an eigenstate of \Jz, and thus the first non-vanishing matrix element is of the fifth order. We remind the reader that this higher-order protection only holds for coherent processes. For incoherent processes, the first interaction excites the atom from the ground state $\ket{\frac52}$ into any eigenstate containing $\ket{\frac32}$, $\ket{\frac52}$ or  $\ket{\frac72}$. From there, a direct path to the other ground state, $\ket{-\frac52}$, might exist depending on the class. We have marked the cases where this distinction matters by an asterisk at the appropriate $J$ values in figures~\ref{fig_c5} and \ref{fig_c_all}.

\textbf{Protection by time-inversion symmetry:} Consider again the two ground states of an atom with integer $J$. As shown above, the two states are connected by time reversal $\T\ket{\Psi_g^+} = e^{i\chi_g}\ket{\Psi_g^-}$. This also implies $\T\ket{\Psi_g^-} = \T\,e^{-i\chi_g}\,\T \ket{\Psi_g^+} = e^{i\chi_g}\T^2\ket{\Psi_g^+} = e^{i\chi_g}\ket{\Psi_g^+}$. 

We now consider one component of $V_1$, $\matx{\Psi_g^-}{\Ji}{\Psi_g^+}$ where $i=x, y \text{ or } z$. Transforming such a matrix element we can write

\begin{align}
\begin{split}
\matx{\Psi_g^-}{\Ji}{\Psi_g^+} &= \matx{\T\Psi_g^+}{e^{i\chi_g}\Ji\,e^{-i\chi_g}\T}{\Psi_g^-}\\
&\qquad\qquad\Big.{\scriptstyle \Ji\T = -\T\Ji } \\
&= -\matx{\T\Psi_g^+}{\T\Ji}{\Psi_g^-}\\
&\qquad\qquad\Big.{\scriptstyle \bracket{\T\alpha}{\T\beta} = \bracket{\beta}{\alpha}} \\
&= -\bracket{\Ji\Psi_g^-}{\Psi_g^+} \\
&\qquad\qquad\Big.{\scriptstyle \bracket{O_H\alpha}{\beta} = \bracket{\alpha}{O_H\beta}} \\
&= -\matx{\Psi_g^-}{\Ji}{\Psi_g^+}\mathperiod
\end{split}\label{timerev}
\end{align}
Thus, for integer $J$ any matrix element of the form $\matx{\Psi_g^-}{\Ji}{\Psi_g^+}$ is zero. E.g.\ the $\pm4\pi/5$ doublet in Fig.~\ref{fig_c5}a for $J > 2$ is protected from first-order transitions. This is the protection mechanism reported by Miyamachi~et~al.\ for Ho atoms on Pt(111)~\cite{Miyamachi2013}.

For half-integer $J$ the argument above does not apply. Instead, time-reversal protection manifests itself in systems where two ground states belong to the same class. These remain magnetic and do not form singlets, since Kramers theorem protects the doublet from mixing and splitting. For those doublets $\Jp\ket{\Psi_g^+}$ and $\Jm\ket{\Psi_g^+}$ both belong to a different class than $\ket{\Psi_g^+}$, so only $\matx{\Psi_g^-}{\Jz}{\Psi_g^+}$ might be non-zero. We have however chosen eigenstates that maximize $\expect\Jz$, setting these matrix elements to zero. Thus also in this case, $V_1$ vanishes. An example of such doublet is the $\pi$ class in Fig.~\ref{fig_c5}c, as indicated in the table Fig.~\ref{fig_c5}d by a yellow background. Note that protection by time reversal only holds for first order electron scattering for both integer and half-integer $J$ systems.

In total, we have identified two different mechanisms that can protect the ground state from zero-energy magnetization flips up to a specific order of electron scattering. First, the states can be protected by time reversal symmetry (yellow background in Fig.~\ref{fig_c5}c \& \ref{fig_c5}d). Second, the states can be protected by rotational symmetry (blue background in Fig.~\ref{fig_c5}c \& \ref{fig_c5}d). The two cases can coexist for certain combinations (green background in Fig.~\ref{fig_c5}c \& \ref{fig_c5}d).

\section{Influence of magnetic field}
Let us now consider how a magnetic field $\vec{B}$ applied to the system affects the protection mechanisms. Strictly speaking, such a field lifts the degeneracy of the doublets, and so transitions within the doublet cannot be excited elastically any more. For finite temperatures, however, the substrate electrons have enough thermal energy to flip the magnetization when $JB \lesssim k_\mathrm{B}T$. Note that he magnetic field may induce other zero-energy transitions, when two states become degenerate due to the Zeeman energy. This is reflected by distinct steps in the hysteresis loop due to Zener tunneling and has been described in detail elsewhere \cite{Wernsdorfer2000}.

In the following, we consider two different situations. First, the external magnetic field is applied along the $z$ direction. Adding a term $\HZ = g_J\Jz B_z$ to the Hamiltonian does not change the rotational symmetry but breaks time reversal symmetry. Thus we expect no changes to the classes of the eigenstates but they will be no longer connected by \T.

For small fields we can treat the Zeeman term as a perturbation. The resulting perturbed ground states $\ket{\Phi_g^\pm}$ can then be written as

\begin{align}
\ket{\Phi_g^\pm} &= \ket{\Psi_g^\pm} + \sum_{\phi^\pm_k = \phi_g^\pm} \frac{\matx{\Psi_k^\pm}{\HZ}{\Psi^\pm_g}}{E_g - E_k}\ket{\Psi^\pm_k} + O(B_z^2)\mathcomma
\end{align}
where $\phi^\pm_k = \phi^\pm_g$ reflects the fact that the matrix element for the Zeeman term is only nonzero if $\Psi_k^\pm$ is in the same class as $\Psi_g^\pm$ \footnote{Note that for half-integer $J$ the other half of the doublet is in the same class and so $E_k$ is zero for one of the summands. However, {\HZ} is diagonal within the doublet subspace and so this term can be neglected.}. As expected, the perturbed states are in the same class as the unperturbed. Thus we only need to consider the breaking of time inversion symmetry, lifting that particular protection. Indeed, if we consider the components of $V_1$ for the new states:

\begin{align}
\begin{split}
\matx{\Phi_g^-}{\Ji}{\Phi_g^+} &= \cancelto{\,=0}{\matx{\Psi_g^-}{\Ji}{\Psi_g^+}}\\
&- 
\sum_{\phi^-_k = \phi_g^-} \frac{\matx{\Psi_g^-}{\HZ}{\Psi^-_k}}{E_l}
\matx{\Psi^-_k}{\Ji}{\Psi_g^+} \\
&-
\sum_{\phi^+_k = \phi_g^+} \frac{\matx{\Psi^+_k}{\HZ}{\Psi_g^+}}{E_k}\matx{\Psi_g^-}{\Ji}{\Psi^+_k}\\
&+ O(B_z^2)\mathperiod
\end{split}
\end{align}

Since $\Psi^+_k$ and $\Psi^-_k$ are connected by time reversal, we can use the same reasoning as in Eqn.~\ref{timerev}, to show that
\begin{align}
\matx{\Psi_a^\pm}{\HZ}{\Psi_b^\pm} &= -e^{i(\chi_a - \chi_b)}\matx{\Psi_b^\mp}{\HZ}{\Psi_a^\mp}\\
\matx{\Psi_a^\pm}{\HZ}{\Psi_b^\mp} &= -(-1)^{2J}e^{i(\chi_a - \chi_b)}\matx{\Psi_b^\pm}{\HZ}{\Psi_a^\mp}
\end{align}
and to rewrite the first sum in terms of $\Psi^+_k$:

\begin{equation}
\begin{split}
\sum_{\phi^-_k = \phi_g^-} \frac{\matx{\Psi_g^-}{\HZ}{\Psi^-_k}}{E_k}
\matx{\Psi^-_k}{\Ji}{\Psi_g^+} = \\
(-1)^{2J}\sum_{\phi^+_k = \phi_g^+} \frac{\matx{\Psi^+_k}{\HZ}{\Psi_g^+}}{E_k}\matx{\Psi_g^-}{\Ji}{\Psi^+_k}\mathperiod
\end{split}
\end{equation}

The resulting first order correction to $V_1$ is thus
\begin{equation}
\left[1 + (-1)^{2J}\right]\sum_{\phi^+_k = \phi_g^+} \frac{\matx{\Psi^+_k}{\HZ}{\Psi_g^+}}{E_k}\matx{\Psi_g^-}{\Ji}{\Psi^+_k}
\end{equation}
and vanishes for half-integer $J$. Note that the second-order terms are a product of three matrix elements, and the corresponding term contains $\left[1 - (-1)^{2J}\right]$, which does not vanish for half-integer $J$ in the general case. Therefore, $V_1$ is proportional to $B_z$ for integer and to $B_z^2$ for half-integer $J$, as noted in yellow boxes in Figures \ref{fig_c5} \& \ref{fig_c_all}. Thus, we expect an increase of flipping rate of the magnetic moment proportional to $B_z^2$ and $B_z^4$ for integer and half-integer spins, respectively. This effect was reported in an STM study for Ho atoms on Pt(111)~\cite{Miyamachi2013}. Note that we can expect the same dependency for any perturbation proportional to \Jz. This includes e.g.\ the diagonal part of the exchange interaction between two nearby atoms 1 and 2 proportional to $\Jz^{(1)}\Jz^{(2)}$. Thus, the lifetime of the magnetization is expected to drop for Ho on Pt(111) when increasing the coverage of Ho adatoms. This is also in agreement with numerical calculations by Hübner et al.~\cite{Huebner2014}, where the magnetic field dependency of the matrix element was computed for the case $\expect{\Jz}=8, q=3$.

For integer $J$ where the crystal field leads to the formation of pairs of singlet states, the magnetic field plays the opposite role. The Zeeman energy reverses the symmetric/antisymmetric pairing due to the crystal field and partially restores magnetic states. $V_1$ is then inversely proportional to $B_z$. For example in the case of Ho with $J=8$ and an $m=\pm8$ ground state in four-fold symmetry, an application of magnetic field results in magnetic states with long lifetimes~\cite{Natterer2017}. This reduces to the situation mentioned above, in which Zener tunneling would induce a step in the hysteresis curve near zero magnetic field. Again, the same effect would be observed for any perturbation proportional to \Jz. As shown elsewhere~\cite{Karlewski2015} also the coupling to the electron bath can undo singlet formation leading to a doublet of ground states with short lifetimes.

In the second case the field is applied perpendicular to $z$ (in-plane). Such a field breaks both symmetries and quenches the magnetic moment along $z$. Only in the case that this field is applied along a different high-symmetry direction (e.g.\ an in-plane two-fold rotation axis in $D_3$) can we expect some remnant magnetization.
For the general case of a tilted magnetic field, no simple statement can be made on the power of the leading order term and effects like cross-over may arise.

\section{Practical symmetries}

\begin{figure*}[!t]
\includegraphics{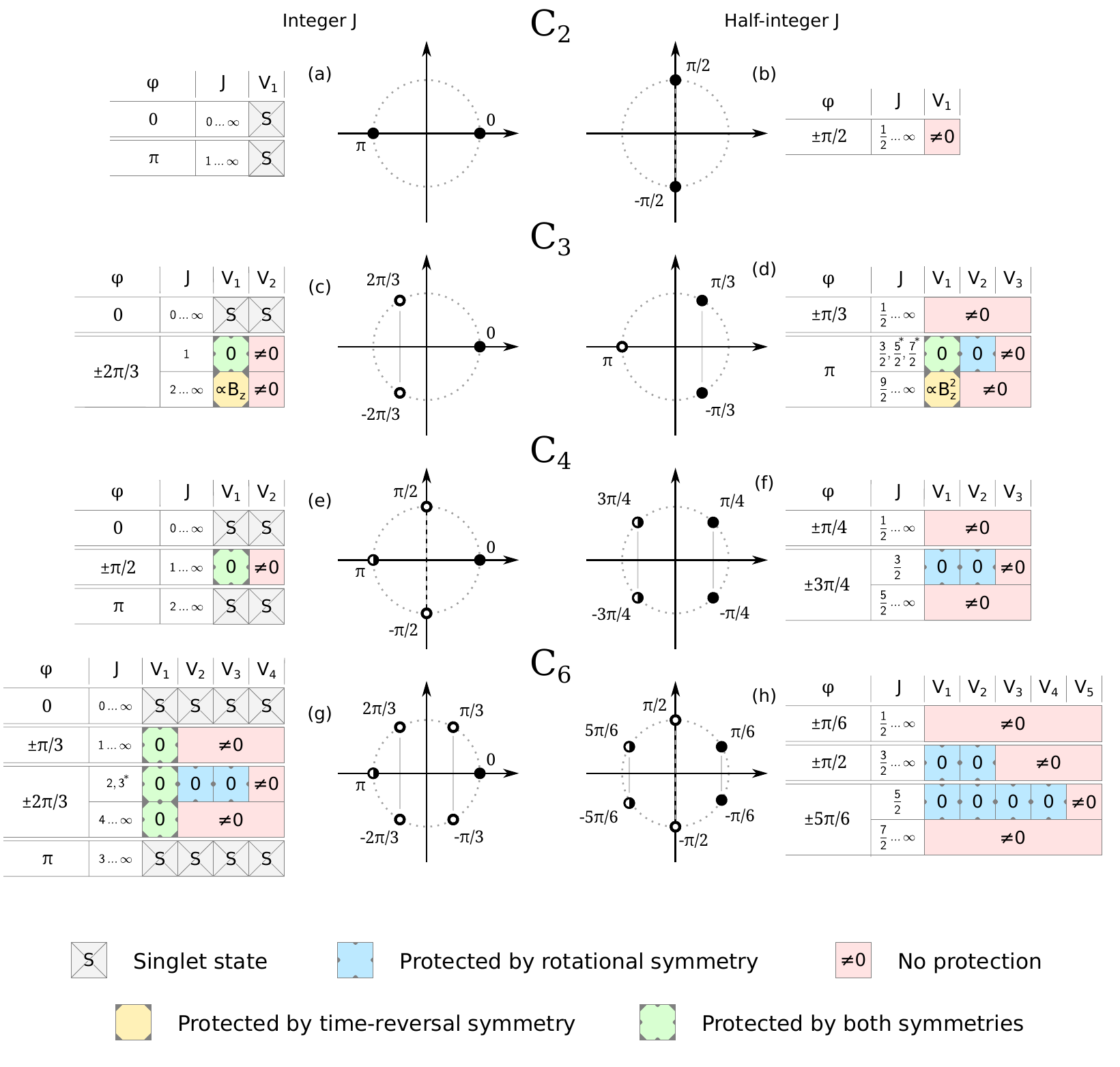}
\caption{Classes and transitions in 2-, 3-, 4- and 6-fold rotation-symmetric environments. Each circle indicates a set of $\ket{m}$ acquiring the same phase under rotation by $2\pi/q$. Empty circles show classes protected against first-order transitions. Half-empty circles indicate classes protected only at low $J$. Asterisks at certain $J$ values indicate that higher order protection is valid only for coherent electron scattering.}
\label{fig_c_all}
\end{figure*}

We can now apply the above analysis to systems of practical value, i.e.\ in 2-, 3-, 4- and 6-fold symmetric environments. The results, which are presented in detail in Fig.~\ref{fig_c_all}, can be summarized as follows:
\begin{itemize}
\item For integer $J$
\begin{itemize}
\item if the two ground states belong to the same class, they are non-magnetic singlets,
\item if the two ground states belong to different classes, at least \textbf{two} electrons are needed for zero-energy magnetization switch.
\end{itemize}
\item For half-integer $J$
\begin{itemize}
\item in 3-fold (6-fold) symmetric environments for the ground state containing $\ket{\pm3/2}$ at least \textbf{two} (\textbf{three}) electrons are needed for zero-energy magnetization switch,
\item otherwise \textbf{one} electron provides for zero-energy magnetization switch.
\end{itemize}
\end{itemize}
For systems with small $J$, higher-order stability can be expected. See Fig.~\ref{fig_c_all} for details.

Finally, for the simple case of rare earth atoms in the oxidation state 3+ and a {\HCF} giving ground states of maximal $\expect{\Jz}$, i.e. the situation best suited for magnetic data storage, we give a simple table that summarizes the findings (Table~\ref{fig_re}). We note that the best candidates for magnetically stable systems are Sm and Ce in 6-fold symmetric adsorption sites.

\begin{table*}
\renewcommand{\arraystretch}{1.5}
\newcolumntype{Y}{>{\centering\arraybackslash}X}
\definecolor{darkblue}{rgb}{0.7,0.9,1.0}
\newcommand{\pr}{\cellcolor{darkblue}}
\definecolor{darkgreen}{rgb}{0.8,1.0,0.8}
\newcommand{\prt}{\cellcolor{darkgreen}}
\definecolor{darkyellow}{rgb}{1.0,0.945,0.717}
\newcommand{\pt}{\cellcolor{darkyellow}}
\begin{tabularx}{0.6\textwidth}{|c|Y|Y|Y|Y|Y|Y|Y|Y|Y|Y|Y|Y|Y|Y|}
\hline
& La & Ce & Pr & Nd & Pm & Sm & Eu & Gd & Tb & Dy & Ho & Er & Tm & Yb\\
$J$ \bigstrut& 0 & $\frac52$ & 4 & $\frac92$ & 4 & $\frac52$ & 0 & $\frac72$ & 6 & $\frac{15}2$ & 8 & $\frac{15}2$ & 6 & $\frac72$\\ \hline
$C_2$ & - & 1 & - & 1 & - & 1 & - & 1 & - & 1 & - & 1 & - & 1 \\ \hline
$C_3$ & - & 1 (2) & \pt 2 & \pt 2 & \pt 2 & 1 (2) & - & 1 (2) & - (2) & \pt 2 & \pt 2 & \pt 2 & - (2) & 1 (2) \\ \hline
$C_4$ & - & 1 & - (2) & 1 & - (2) & 1 & - & 1 & - (2) & 1 & - (2) & 1 & - (2) & 1 \\ \hline
$C_6$ & - & \pr 5 & \prt 2 & \pr 3 & \prt 2 & \pr 5 & - & 1 (3) & - (2) & \pr 3 & \prt 2 & \pr 3 & - (2) & 1 (3) \\ \hline
\end{tabularx}
\caption{Electron-assisted magnetization tunneling in single 3+ rare-earth ions under the assumption of ground states with maximum $\expect{\Jz}$. The numbers denote the smallest number of electron scattering processes needed for a transition between the two ground states (lowest $k$ such that $V^k \neq 0$). ``-'' denotes singlet ground states. Numbers in parentheses indicate maximum possible stability for arbitrary ground states.}
\label{fig_re}
\end{table*}

\section{Higher symmetries}

We now analyze the stability of degenerate doublets in other axial symmetries with a $q$-fold rotation axis. We show that these symmetries behave either like $C_q$ or like $C_{2q}$. In the latter case the stability is greatly enhanced --- e.g.~in a $D_{5d}$-symmetric environment there are 10 classes, so that ground states in a magnetic atom with integer $J$ are protected up to second order in most of the cases.

\subsection{$C_{qv}$ and $D_q$}
$C_{qv}$ adds a vertical mirror plane $\sigma_v$ to $C_q$. The $\sigma_v$ operator anticommutes with {\Jz} and does not commute with \R: $\sigma_v\R = \R[-1]\sigma_v$. Thus, the eigenstates we chose will not be eigenstates of $\sigma_v$, and their composition will be the same as for $C_q$. Marciani et al.~have shown that the presence of the mirror plane introduces only a weak constraint on the switching process~\cite{Marciani2017}.

$D_q$ adds a 2-fold rotation axis perpendicular to the main axis to $C_q$. The {\Rx} operator behaves exactly like $\sigma_v$ with respect to {\R} and the components of $\vec{J}$. Thus, again, the stability in $D_q$ will follow the same rules as in $C_q$.

\subsection{$C_{qh}$ and $D_{qh}$}
The rules change in the environments with a horizontal mirror plane $\sigma_h$. The $\sigma_h$ operator commutes with {\Jz} and anticommutes with \J{x}, \J{y} and \J{\pm}. Thus $\ket{m}$ are eigenstates of $\sigma_h$ with eigenvalue sign alternating between neighboring $m$'s. We can write $\sigma_h\ket{m} = (-1)^{J+m}\ket{m}$. $\sigma_h$ also commutes with \R, and thus the eigenstates of {\HCF} can be chosen to be eigenstates of both {\R} and $\sigma_h$. In fact, the eigenstates we chose in section~\ref{symmetries} are already eigenstates of both operators. This is easy to see in the case of doublets where the two states belong to different classes, as there the choice is unique. For same-class doublets we note that they only exist for half-integer $J$ and odd $q$ and are composed of kets with $m = q(n + \frac12)$. If such a state is an eigenstate of $\sigma_h$, then for every $n$ it can contain either $\ket{m}$ or $\ket{-m}$, but not both, since $\matx{-m}{\sigma_h}{-m} = (-1)^{2m}\matx{m}{\sigma_h}{m} = (-1)^q\matx{m}{\sigma_h}{m} = -\matx{m}{\sigma_h}{m}$. It follows that for every $n$ one of the doublet's states will contain $\ket{m}$, while the other $\ket{-m}$, and so the matrix element of {\Jz} between these states will be zero, consistent with out choice from section~\ref{symmetries}.

Now every state is characterized by the rotational phase $\phi$ and the eigenvalue of $\sigma_h$ ($\pm1$). Note that for even $q$ all the states with the same $\phi$ gain the same sign after reflection. Thus for even $q$ the number of classes does not change, and the same analysis applies. For odd $q$, however, half of the $\ket{m}$ kets gain one sign and the other half the other, splitting every class in two. In this case we can reclassify the states according to a new phase $\phi_{2q} = \frac{\pi m}q$, that would normally arise in $C_{2qv}$ symmetry. The new classes combine states with a same eigenvalues of \R and $\sigma_h$ and have the same properties we have used before, namely:
\begin{enumerate}
\item $\J{\pm}$ operators switch the class one step forward or backward along the unit circle,
\item {\Jz} operator does not affect the class,
\item {\T} connects classes with opposite sign of $\phi_{2q}$.
\end{enumerate}
Thus the stability rules in $C_{qh}$ are the same as the stability rules in $C_q$ for even $q$ and $C_{2q}$ for odd $q$. The $D_{qh}$ behaves the same way, since the additional {\Rx} symmetry does not have a strong effect on stability.

\subsection{$D_{qd}$ and $S_{2q}$}
The $S_{2q}$ point group is generated by the rotation-reflection operator $\Sq = \Rhalf\sigma_h$ with $\Sq\Sq = \R$. {\Sq} commutes with {\Jz} and anticommutes with \J{x}, \J{y} and \J{\pm}. $\Sq\ket{m} = (-1)^{J+m}e^{i\pi m/q}\ket{m}$. The choice of eigenstates as eigenstates of both {\Ham} and $\Sq$ again maximizes $\expect{Jz}$ for every doublet. We can thus use the phase $\phi_s = \pi m (1 + 1/q)$ to classify the states. In the case of even $q$ we obtain $2q$ classes, while in the case of odd $q$ there are only $q$ classes. Based on this, we can reclassify the states again based either on $\phi_{2q}$ or $\phi$, depending on the parity of $q$. As in the previous case, this preserves the composition of classes and the action of \Jz, \Jp, {\Jm} and {\T} with respect to the unit circle of classes. It follows that the stability rules in $S_{2q}$ are the same as the stability rules in $C_q$ for odd $q$ and $C_{2q}$ for even $q$. The $D_{qd}$ has additionally a vertical mirror plane, that does not alter stability rules.

Table \ref{similar} summarizes the found relations of higher order symmetries to the behavior of the simpler $C_q$ symmetry.

\begin{table}
\renewcommand{\arraystretch}{1.5}
\begin{tabular}{c||c|c||c|c|c||c|}
\hline
$q$ & $\ C_{qv}\ $ & $\ C_{qh}\ $ & $\ D_{q}\ $ & $\ D_{qh}\ $& $\ D_{qd}\ $ & $\ S_{2q}\ $ \\ \hline
odd & \multirow{2}{*}{$C_q$} & $C_{2q}$  & \multirow{2}{*}{$C_q$} & $C_{2q}$ & $C_q$ & $C_q$ \\ \cline{1-1} \cline{3-3} \cline{5-7}
even & & $C_q$  & & $C_q$ & $C_{2q}$ & $C_{2q}$ \\ \hline
\end{tabular}
\caption{Similarity regarding magnetic stability between axial point groups. An eigenstate in any of the listed groups behaves as it would in the corresponding $C_q$ or $C_{2q}$ group.} \label{similar}
\end{table}

\section{Magnetic stability of H\lowercase{o} atoms}

There have been several reports of stable and unstable single rare earth atom systems, mostly based on Holmium single atoms. Here we review their results in the light of our stability criteria.

Ho atoms on three-fold symmetric adsorption sites on Pt(111) have been investigated by STM~\cite{Miyamachi2013} and XMCD~\cite{Donati2014}. Donati et al. report a ground state with $J=8, m=\pm6$ (class $0$), obtained by fitting XMCD spectra based on multiplet calculations. In this configuration there are two singlet states, that will mix in a high magnetic field, leading to a hysteresis-free magnetization curve, as measured in the experiment. Ab initio calculations from the paper by Miyamachi et al., on the other hand, predict a ground state with $J=8, m=\pm8$ (classes $\pm2\pi/3$). Such ground states are protected by time reversal symmetry, so we expect long lifetimes, unless a magnetic field is applied. The STM experiments showed long lifetimes at zero field that become shorter in small magnetic fields. The energy of an inelastic excitation observed in the experiment, also agrees with the theoretically predicted $J$ multiplet structure.

Interestingly, an ensemble of Ho atoms with time-reversal-protected ground state doublet would also show no magnetic hysteresis under the experimental conditions of the XMCD measurements ($T=\SI{2.5}{\kelvin}$, $\approx\SI{10}{\second}$ per point). At zero field the average magnetization of Ho atoms is zero, as the two stable magnetization orientations are equally probable. The application of a magnetic field along $z$ splits the doublet, introducing a preferential orientation, while at the same time increasing the switching rate quadratically, leading to an overall increase in average magnetization. At high fields the two states split so far that switching rate becomes negligible and the magnetization saturates. As the magnetic field is reduced, the switching rate increases and then decreases again for vanishing magnetic field. Since the energy difference between the two states also goes down to zero, we again expect equal population of the ground states at zero field and no hysteresis. The stability at vanishing magnetic fields would be also influenced by the distance between Ho atoms due to exchange or dipolar interactions. As has been shown above, any operator linear in $J$ acts like a magnetic field. Note that the STM experiments have been carried out at extremely low coverage~\cite{Miyamachi2013}, while the XMCD data was taken at a coverage of \SI{0.04}{\percent}~\cite{Donati2014}.

Thus the STM and the XMCD results agree on the absence of hysteresis, but disagree on the multiplet structure. The question about the configuration of the crystal field and the ground state of this system remains open. Shick et al.~\cite{Shick2017} have proposed that the crystal field parameters and thus $\expect{\Jz}$ of the ground state changes with magnetic field, due to an interaction between $5d$ and $4f$ orbitals, such that in low magnetic fields one expects $\expect{\Jz} \approx 8$, and in high magnetic fields $\expect{\Jz} \approx 6$.

Similar to Ho on Pt(111), STM experiments~\cite{Natterer2017} and XMCD experiments~\cite{Donati2016} on Ho atoms in 4-fold symmetric adsorption sites on MgO fundamentally disagree on the class of the ground state. The ground state deduced from XMCD spectra is the class $\pm\pi/2$ doublet ($\expect\Jz=4.66$). It is protected from switching by rotation symmetry and so the magnetization curves show a hysteresis. The estimated crystal field parameters predict the first excited state \SI{4.5}{\milli\electronvolt} higher than the ground state. At the same time, the STM experiments report the first excited state at \SI{73}{\milli\electronvolt} above the ground state. Natterer et al.\ estimate $m=\pm8$ (class $0$) from the Ho magnetic moment. This ground state should be non-magnetic, but as the experiments are done in a mostly in-plane magnetic field, the two singlet states would mix and acquire a magnetic moment. Transitions between such mixed states are, in general, allowed, but the actual transition probability depends on the strength of the crystal field. In the STM setup the magnetic state was stable for several hours, likely indicating weak transversal anisotropy and, as expected, weak interaction with the electron bath. We would expect little or no hysteresis in this system, as at zero field the magnetic states should lose their magnetization.

As in the previous case, the origin of the disagreement between STM and XMCD data on the energy spectrum in Ho on MgO remains open and needs further investigation. Possible explanations such as different MgO thickness and averaging over several adsorption sites in the XMCD experiment can only partially explain the discrepancies.

\section{Comparison to other theoretical results}

\subsection{M.~Marciani et al. (2017)}

The paper by Marciani et al.~\cite{Marciani2017} uses an approach very similar to ours. However, we disagree with several points of the analysis. First, in the case of $q=3$ Marciani et al.\ write that for $J=9/2$ the ground state is not stable, based on their analysis of the structure of the Hamiltonian. This is not confirmed by a direct construction of an appropriate Hamiltonian and calculation of the transition rates. The relevant block of the Hamiltonian in the basis $\ket{-\frac92}, \ket{-\frac32}, \ket{\frac32}, \ket{\frac92}$ has a general form
$$
\begin{bmatrix}
2d & a & b & 0\\
\bar{a} & 0 & 0 & b\\
\bar{b} & 0 & 0 & -a\\
 0 & \bar{b} & -\bar{a} & 2d
\end{bmatrix},
$$
plus a constant term, and can be diagonalized analytically. The eigenstates with maximal $\expect\Jz$ within every doublet are not connected by either \Jp, {\Jm} or \Jz. Thus first-order electron transitions are forbidden between these states.

Second, Marciani et al.\ assume large uniaxial anisotropy, such that $\expect\Jz$ of the ground state is very close to $J$. For the case where the ground state has a smaller $\expect\Jz$, they propose using a smaller $J_\mathrm{eff}\approx\expect\Jz$ for analysis of the ground state stability. This, however, might lead to wrong conclusions if $2J \geq q$ while $2J_\mathrm{eff} < q$, as the latter suggests stable states. Consider, for example, $J=\frac72$ in $C_6$, and two doublets from the $\phi=\pm\frac56\pi$ class pair, one with $\expect{\Jz} \approx \frac72$ and the other with $\expect{\Jz} \approx \frac52$. Both are unstable, as they contain states from neighboring classes. However, if the second doublet has lower energy, Marciani et al. suggest using $J_\mathrm{eff} = \frac52$ to determine the stability. But in a system with $J=\frac52$ the eigenstates of {\HCF} are eigenstates of \Jz, and the doublet $\expect{\Jz}=\frac52$ is stable up to fourth order. Using this conclusion for the $\expect{\Jz} \approx \frac52$ doublet in $J=\frac72$ system is definitely unwarranted.

To avoid such mistakes, one could amend the definition of $J_\mathrm{eff}$ to include the class of the doublet in question instead of its $\expect{\Jz}$. That is, one should choose $J_\mathrm{eff}$ equal to the highest $m$ that belongs to the same class. For the example above that would be $m = \frac72$ for both doublets, and thus the same conclusion on stability will be reached. One should keep in mind, however, that such a definition would only work for single-electron processes, as higher-order stability also depends on the neighboring classes. As an example, consider the $\pm\frac23\pi$ doublet for $J=2$ and $3$ in $C_6$. $J_\mathrm{eff}$ is equal to 2 for both cases, but for $J=3$ second-order transitions are possible at non-zero temperature, while for $J=2$ they are forbidden.

\subsection{M.~Prada (2017)}

The paper by M.~Prada~\cite{Prada2017} classifies single atom systems by introducing a geometric phase, as a phase difference between two equivalent paths connecting the two ground states. However, this phase was introduced based on the equality $\T\R = \R[-1]\T$, which, due to commutativity~\cite{bookGT, bookQM, Marciani2017} of {\T} and {\R} only holds for $q=2$. The operators $\T^+$ and $\T^-$ are thus not equivalent to \T, as the paper claims, but to \R[2]\T\ and \R[-2]\T, and the proposed phase is essentially double the eigenvalue of {\R} for the state. This has reprecussions throughout the paper, as for example, $\T^+$ and $\T^-$ do not anticommute with \J{x}\ or \J{y}, and thus Eqn.~15 in Ref.~\onlinecite{Prada2017} derived under the assumption $\T^+\vec{J} = -\vec{J}\T^+$ becomes invalid. Through these mistakes the final classification scheme~\cite{Prada2017} gives wrong results in almost one half of the cases: for half-integer $J$ all doublets in 3-, 4- and 6-fold symmetric environments are listed as protected. A direct calculation shows that this is untrue. For example, let us consider the case of $J=\frac72$ in $C_{3v}$ and ground states from the $\phi=\pm\frac\pi3$ class pair. The two relevant blocks of the Hamiltonian in $\ket{-\frac72}, \ket{-\frac12}, \ket{\frac52}$ and $\ket{-\frac52}, \ket{\frac12}, \ket{\frac72}$ bases have a general form
$$
\begin{bmatrix}
e_1 & a & c\\
\bar{a} & e_2 & b\\
\bar{c} & \bar{b} & e_3
\end{bmatrix} \text{ and }
\begin{bmatrix}
e_3 & -b & c\\
-\bar{b} & e_2 & -a\\
\bar{c} & -\bar{a} & e_1
\end{bmatrix}.
$$ We could now calculate the $V_1$ analytically, but since it is enough to show that $V_1$ is not zero for specific values of the constants, we choose to calculate matrix elements $\matx{\Phi_g^-}{\J{\pm}}{\Phi_g^+}$ numerically for $e_1=-1$, $e_2=0$, $e_3=-0.5$, $a=b=0.01$, $c=0.0001$. These values satisfy the criteria of $\expect{\Jz} \approx J$, out-of-plane easy axis and small transversal terms. We get a ground state energy of approximately -1.0001 and $\matx{\Phi_g^-}{\Jm}{\Phi_g^+} = 0.0004$, $\matx{\Phi_g^-}{\Jp}{\Phi_g^+} = 0$.

Based on our analysis of the problem, we believe that it is impossible to classify all the possible systems based on a single number. However, if we ignore the complications arising at small $J$ and in magnetic field, and focus on first-order interactions, it becomes possible. First, we note that for integer $J$ everything except single states is protected, and single states gain a phase $\phi_g = 0\text{ or }\pi$ under rotation by \R. Thus we write that the doublet is stable unless $\cos^2{\phi_g} = 1$. For half-integer $J$ the doublet is stable, unless the two states belong to neighbouring classes. This we write as $\cos^2{\phi_g} = \cos^2\frac{\pi}q$. We combine the two cases and introduce a number $\eta = \cos^2{\phi_g} - \cos^2{(\pi J)} - \cos^2\frac{\pi}q\sin^2{(\pi J)}$. Now given an eigenstate that belongs to a rotation class with phase $\phi_g$, we conclude that this state belongs to a magnetic doublet, protected in first order against magnetization tunneling, unless $\eta = 0$.

To additionally check the correctness of our results, numerical calculations have been performed for $q=2,3,4,5\text{ and }6$ and $J$ up to 12 using Hamiltonians based on Stevens' operators~\cite{Stevens1952}. The corresponding \textsf{python} code is made available on Github~\cite{github}.

\section{Conclusions}

We have analyzed the stability of magnetization in arbitrary doublets in 2-, 3-, 4- and 6-fold symmetric adsorption sites. We show that depending on the symmetry, the value of $J$ and the ground state doublet, different order processes are needed for a zero-energy magnetization switch. The most stable doublet requires a 5th order scattering process to switch ($C_6: J=5/2, m=5/2$).

In real life systems the protection will not be absolute. Presence of magnetic fields (particularly in-plane) and Jahn-Teller distortion can break the symmetries that protect the magnetization. Further, at $T > \SI{0}{\kelvin}$ also phonons can induce transitions, as they distort the lattice, temporarily inducing mixing of the states. This breaks both protection mechanisms. Similarly, the hyperfine interaction may lead to additional constraints on magnetization tunneling that depend on magnetic field. However, in the limit of low fields and a thermal population of the nuclear spin states, the nuclear spins may be treated as a simple heat bath in analogy to the electron heat bath. When taking the lowest order interaction of the electronic total angular momentum $J$ with the nuclear spin $I$ of the form $A\vec{I}\vec{J}$, the selection rules are identical to electron scattering not qualitatively changing magnetic stability. Only, if simultaneous spin flips of a conduction electron and a nuclear spin are considered, the rules for elastic magnetization reversal change and stability can only be sustained for protection of at least second order.
The difference between a theoretically stable and theoretically unstable system can however make a decisive difference in the lifetime of the magnetic state.

The considerations above would limit the lifetime of any system. Magnetic stability cannot be achieved over arbitrary times. For applications of single atoms on surfaces for quantum information processing, additionally the decoherence time $T_2$ is of importance. A functioning device must operate on the quantum states faster than the decoherence time and the magnetization lifetime. We note that maximally stable states (long lifetimes) are also maximally classical. If one prepares a superposition of the two ground states, e.g. $\frac1{\sqrt2}\left(\ket{\Psi_g^+} + \ket{\Psi_g^-}\right)$, it only takes the elastic interaction with a single electron of the bath to completely dephase the superposition.

\section{Acknowledgement}

We acknowledge financial support by the German Research Foundation under the grant WU 349/4-2.

\bibliography{References}

\end{document}